%% file: icme25.tex
\documentclass[conference]{IEEEtran}
\IEEEoverridecommandlockouts

\usepackage{algorithm}
\usepackage{algorithmic}
\usepackage{url}          
\usepackage{booktabs}      
\usepackage{amsfonts}      
\usepackage{nicefrac}     
\usepackage{microtype}     
\usepackage{xcolor}        
\usepackage{xspace}
\usepackage{multirow}

\usepackage{amsmath}
\usepackage{amssymb}
\usepackage{mathtools}
\usepackage{amsthm}
\usepackage{multirow}
\usepackage{algorithm}
\usepackage{algorithmic}
\usepackage{subfigure}
\usepackage{cite}
\usepackage{amsmath,amssymb,amsfonts}
\usepackage{algorithmic}
\usepackage{graphicx}
\usepackage{textcomp}
\usepackage{xcolor}
\newcommand{\ie}{{\em i.e.},\xspace}
\newcommand{\eg}{{\em e.g.},\xspace}

\def\BibTeX{{\rm B\kern-.05em{\sc i\kern-.025em b}\kern-.08em
    T\kern-.1667em\lower.7ex\hbox{E}\kern-.125emX}}

\begin{document}

\title{Towards Imperceptible Adversarial Attacks for Time Series Classification with Local Perturbations and Frequency Analysis}

\author{
  \IEEEauthorblockN{
    Wenwei Gu\IEEEauthorrefmark{1},
    Renyi Zhong\IEEEauthorrefmark{1},
    Jianping Zhang\IEEEauthorrefmark{1}\IEEEauthorrefmark{2}\thanks{\IEEEauthorrefmark{2} Jianping Zhang is the corresponding author.},
    Michael R. Lyu\IEEEauthorrefmark{1}
  }
  \IEEEauthorblockA{\IEEEauthorrefmark{1}The Chinese University of Hong Kong, China.\\
    Email: \{wwgu21, ryzhong22, jpzhang, lyu\}@cse.cuhk.edu.hk}
}

\maketitle

\begin{abstract}
Adversarial attacks in time series classification (TSC) models have recently gained attention due to their potential to compromise model robustness. Imperceptibility is crucial, as adversarial examples detected by the human vision system (HVS) can render attacks ineffective. Many existing methods fail to produce high-quality imperceptible examples, often generating perturbations with more perceptible low-frequency components, like square waves, and global perturbations that reduce stealthiness. This paper aims to improve the imperceptibility of adversarial attacks on TSC models by addressing frequency components and time series locality. We propose the Shapelet-based Frequency-domain Attack (SFAttack), which uses local perturbations focused on time series shapelets to enhance discriminative information and stealthiness. Additionally, we introduce a low-frequency constraint to confine perturbations to high-frequency components, enhancing imperceptibility. Comprehensive experiments demonstrate our method’s superiority, achieving comparable performance to state-of-the-art baselines while reducing perturbation magnitude by about 17.4\%.

\end{abstract}


\input{contents/01_introduction}
\input{contents/02_related_works}
\input{contents/03_methodology}
\input{contents/04_experiments}
\input{contents/05_conclusion}

\bibliographystyle{IEEEbib}
\bibliography{icme25}

\input{contents/06_appendix}

\end{document}

%% file: contents/01_introduction.tex
\section{Introduction}

Time series analysis has become one of the central topics in the field of modern data mining with the growing volume of temporal data collected from sensors ~\cite{foumani2023deep, ismail2019deep}, while time series classification (TSC) is one of the key time series analysis task~\cite{esling2012time}. There are many applications of time series classification, including human activity recognition~\cite{chen2021deep}, ECG-based patient diagnosis~\cite{schirrmeister2017deep} and audio identification~\cite{yang2021voice2series}. With the advent of deep learning, neural network models have demonstrated superior performance in various tasks, including time series classification~\cite{ding2023black}. However, deep neural networks (DNNs) are vulnerable to adversarial attacks~\cite{szegedy2013intriguing, goodfellow2014explaining}, where small perturbations are added to deceive the deep learning model to make mispredictions. There are several significant security concerns, \eg the used classification algorithms in healthcare devices can be tricked into misdiagnosing patients, which can affect the diagnosis of their disease~\cite{karim2020adversarial}. Thus, several studies have been conducted on adversarial attacks and defenses on deep learning-based time series classification~\cite{ismail2019deep, pialla2022smooth, jiang2020adversarial}.

Various adversarial attack methods have been proposed for image classification~\cite{zhang2024improving}. For image classification, an adversarial attack consists of modifying several pixels in the original image such that the changes are almost undetectable by a human~\cite{yuan2019adversarial}. For adversarial attacks on time series, there are some main differences compared to those on images, which lie in the different visualization of both modalities of data and the different perceptual sensitivities to colors and time series curves of human beings~\cite{pialla2022smooth}. In other words, slightly changing a few pixels will make the images look nearly the same and not affect how the human visual system (HVS) perceives the image. On the contrary, HVS can easily discriminate the adversarial time series as the curve changes of the perturbed time series are distinct. Indeed, perturbation on a time series can be analogous to perturbation on a row or column of pixels of the image. Thus, compared with attacks on images, generating imperceptible adversarial examples for TSC with a high attack success rate (ASR) can be more challenging. 

\begin{figure*}
  \centering
  \subfigure[Nearly square wave perturbation by FGSM]{
  \label{FGSM Example}
  \includegraphics[width=0.93\columnwidth]{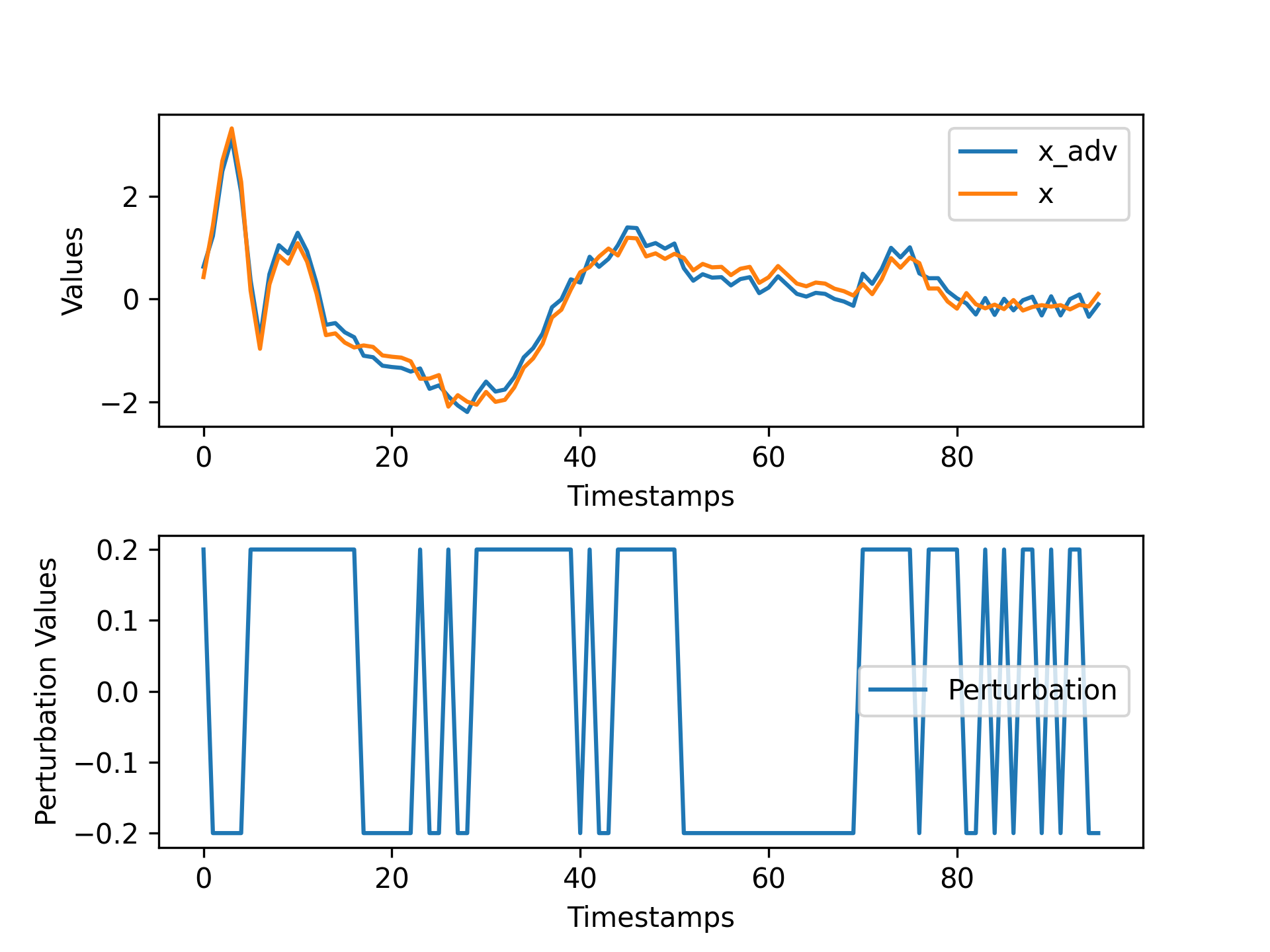}
  }
  \subfigure[Perturbation with less low frequency component by SFAttack]{
  \label{SFAttack Example}
  \includegraphics[width=0.93\columnwidth]{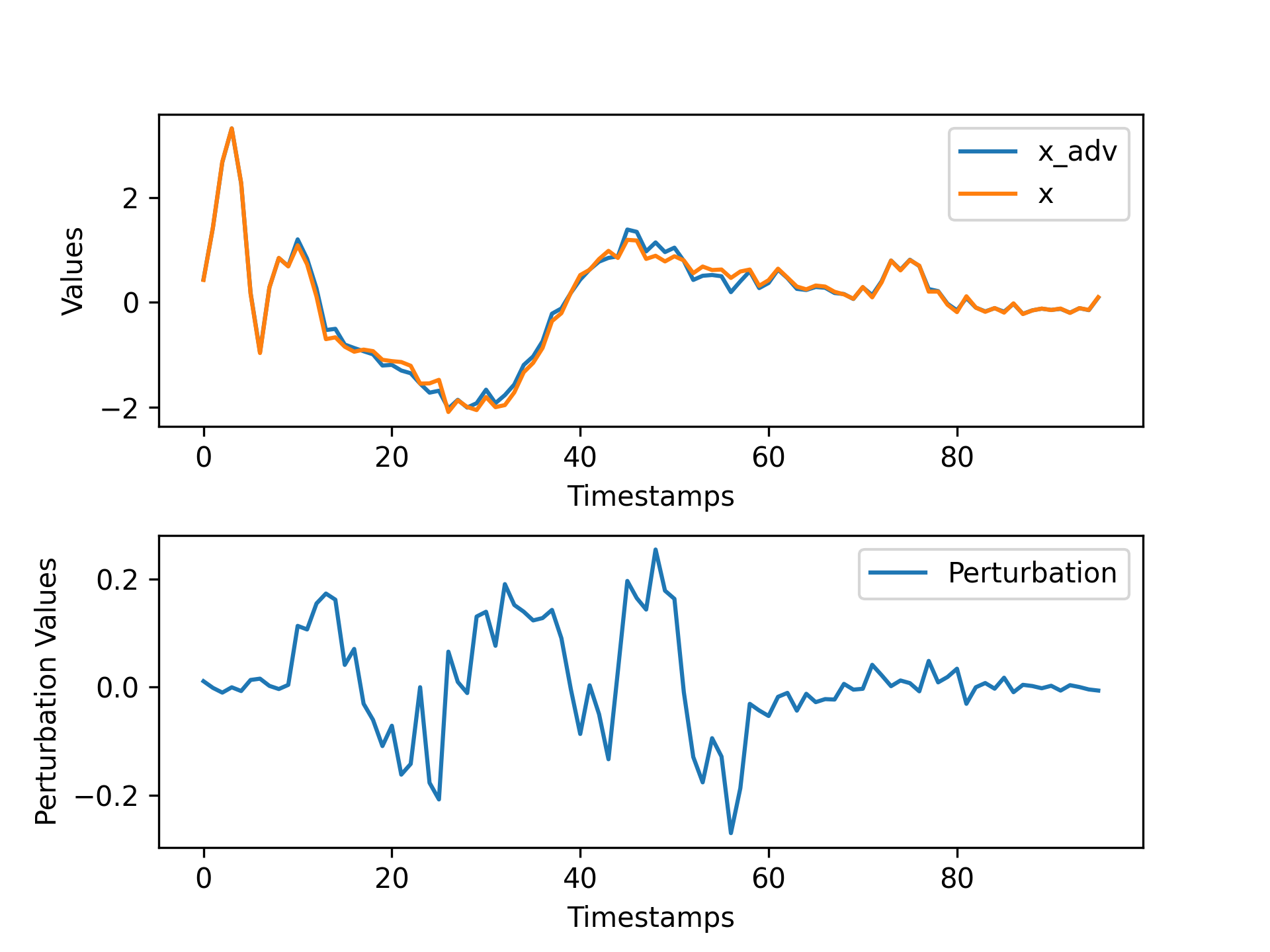}
  }
  \vspace{-1mm}
  \caption{Adversarial examples generated by FGSM and SFAttack in EGG dataset}
  \vspace{-2mm}
  \label{Examples}
\end{figure*}

Existing works on adversarial attacks on TSC models typically focus on constructing effective adversarial examples that increase the misclassification rate. For instance, in~\cite{ismail2019deep}, the fast gradient sign method (FGSM) is adopted to create adversarial examples. Though achieving high performance, these methods suffer from low imperceptibility due to two reasons, leading to easily identifiable adversarial examples~\cite{belkhouja2020analyzing}. On one hand, these adversarial examples are obtained through global perturbation in the whole time series~\cite{wu2022small}, however, global perturbation can be more easily perceived compared to local perturbations. On the other hand, existing methods typically produce adversarial examples through time-domain attacks. However, the perturbation obtained through these methods contains some low-frequency components, \ie square waves, with abrupt level shifts (up or down), which are easily perceived by HVS as shown in Figure~\ref{FGSM Example}. Their overlook of crucial frequency information in time series can significantly undermine the imperceptibility. 

In this paper, we propose the Shapelet-based Frequency-domain Attack (SFAttack) to overcome the aforementioned limitations of existing approaches. To improve imperceptibility, SFAttack attempts to conduct local perturbations mainly on the discriminative segments of time series, namely, the time series shapelets~\cite{ye2009time}. This is based on the intuition that perturbing the most representative patterns of time series can make it more deceptive. Furthermore, inspired by~\cite{luo2022frequency, guo2020low,long2022frequency} that works for image classification attack, we introduce the discrete cosine transform (DCT) to limit the low-frequency component of the perturbation. This is because low-frequency perturbations have worse imperceptibility for several attacking defended models~\cite{sharma2019effectiveness}, \ie an elimination on the low-frequency components can further enhance the imperceptibility. 

We conduct extensive experiments to validate the effectiveness of our proposed SFAttack on three widely adopted time series classifiers. Remarkably, our SFAttack can achieve comparable attack success rates with higher adversarial imperceptibility measured by the reduction of $L_2$ loss at around 17.4\%. Moreover, we have also provided some theoretical analysis that motivates our design.

We summarize the main contributions as follows: 

\begin{itemize}
    \item {To improve the imperceptibility of the adversarial time series, we propose the Shapelet-based Frequency-domain Attack (SFAttack). SFAttack utilizes the time series shapelet to conduct local perturbation to limit the perturbation area. It also leverages the strength of frequency domain transforms to eliminate the impact of the low-frequency components of the perturbation that further enhance the imperceptibility.}
    \item {We conduct the theoretical analysis of the design of SFAttack, namely, the motivation of using $L_2$ norm as the constraint and the strength of incorporating time shapelets.}
    \item {We conduct comprehensive experiments to validate the advantages of SFAttack on three widely adopted time series classifiers, observing that the imperceptibility of adversarial samples is largely enhanced while the performance remains comparable to state-of-the-art methods.}
\end{itemize}

%% file: contents/02_related_works.tex
\section{Related Works}

\subsection{Time Series Classification}

The target of TSC is to learn discriminative patterns that can be used to classify time series into predefined categories. With the development of deep learning techniques, several deep learning-based approaches have achieved superior performance on TSC. The temporal convolutional network (TCN)~\cite{koh2021deep} is one of the representative methods that combine dilation and residual connections with causal convolutions. ResCNN~\cite{zou2019integration} is a residual network with CNN for TSC. In ResCNN, the fully connected layer is replaced by a global average pooling in order to alleviate overfitting. As one of the state-of-the-art in the TSC task, InceptionTime~\cite{ismail2020inceptiontime} is an ensemble of five Inception-based convolutional neural networks. Besides deep learning methods, Rocket~\cite{dempster2020rocket} is an efficient TSC algorithm that transforms the input time series using random convolutional kernels followed by the proportion of positive values and max pooling operators. The transformed features are then utilized to train a linear classifier. MiniRocket~\cite{dempster2021minirocket} is a faster variant of Rocket that uses a small and fixed set of kernels, which is the default variant in the Rocket family. 

\subsection{Adversarial Attacks on TSC}

Existing TSC models have achieved high performance, however, they are vulnerable to adversarial attacks by adding small perturbations to the input time series~\cite{ding2023black}. Specifically, traditional adversarial attack methods on image classification like FGSM~\cite{goodfellow2014explaining} and BIM~\cite{kurakin2016adversarial} have been introduced to TSC in \cite{ismail2019deep}. FGM is a variant of FGSM that computes the perturbation under the $L_2$ norm constraint rather than $L_\infty$ norm constraint~\cite{dong2018boosting}. PGD~\cite{madry2018towards} is an extension of FGM that performs multiple iterations of perturbation to generate adversarial examples. C\&W~\cite{carlini2017towards} utilize the distance between the original and adversarial example as a regularization item to the objective function. More recently, a specifically designed method for TSC, SGM~\cite{pialla2022smooth} adds fuzzed lasso regularization to smooth the generated perturbations. 

Imperceptibility is a crucial requirement for adversarial attacks~\cite{zhang2024curvature, qin2019imperceptible}, which ensures the generation of high-quality adversarial examples that evade the detection by HVS. Nevertheless, the concept of imperceptibility varies significantly between images and time series due to their inherent differences in visualization. For instance, an attack method that perturbs only a single pixel within an image has been described in~\cite{su2019one}. Such a minor perturbation is almost imperceptible due to the high-dimensional nature of images. However, in the context of time series, perturbations that modify only a single data point can be remarkably noticeable. Intuitively, perturbing a point in a time series can be likened to perturbing an entire line in an image, given the sequential structure of time series data. This discrepancy in imperceptibility between images and time series underscores the need for a deeper exploration into the imperceptibility of time series data. In our solution, we resort to local perturbations that focus on the discriminative shapelets and the frequency transform that eliminates the low-frequency components of the time series.

%% file: contents/03_methodology.tex
\section{Methodology}

\subsection{Preliminary}

A time series is a sequence vector $x\in{R^T}$ , $x=[x_1, ..., x_T]$, with $c$ as its ground truth label of class and $T$ as its length. Given a time series classifier $f(\cdot)$ and the time series $x$, an adversarial attack aims to perturb the classifier by adding a small variation $p$ to $x$. The $p$ is referred to as noise or perturbation. The perturbed time series $x_{adv} = x + p$ is called an adversarial sample. The attack is considered successful if the class predicted for the adversarial sample is different from the groundtruth label of the original time series, \ie $f(x_{adv}) \neq c$. With the aim of imperceptibility, the added noise $p$ should be small enough to satisfy the $L_q$ constraint. Specifically, the predefined perturbation budget $\epsilon$ should satisfy $\|p|_q\leq\epsilon$. 

\subsection{Overview}

The goal of our Shapelet-based Frequency-domain Attack (SFAttack) is to improve adversarial imperceptibility. Firstly, we propose to use the time series shapelets to identify critical areas for performing local perturbations, as a more focused solution that utilizes the characteristic of time series. Secondly, we have introduced the use of Discrete Cosine Transform (DCT) to carry out attacks in the frequency domain, effectively eliminating the HVS perceptible low-frequency components. Furthermore, we have applied the FGM-based optimizer and $L_2$ norm constraint, which further enhances the stealthiness and effectiveness of our attacks.

\subsubsection{Shapelet-Based Attack}

\begin{figure}
\centering
\includegraphics[width=.9\linewidth]{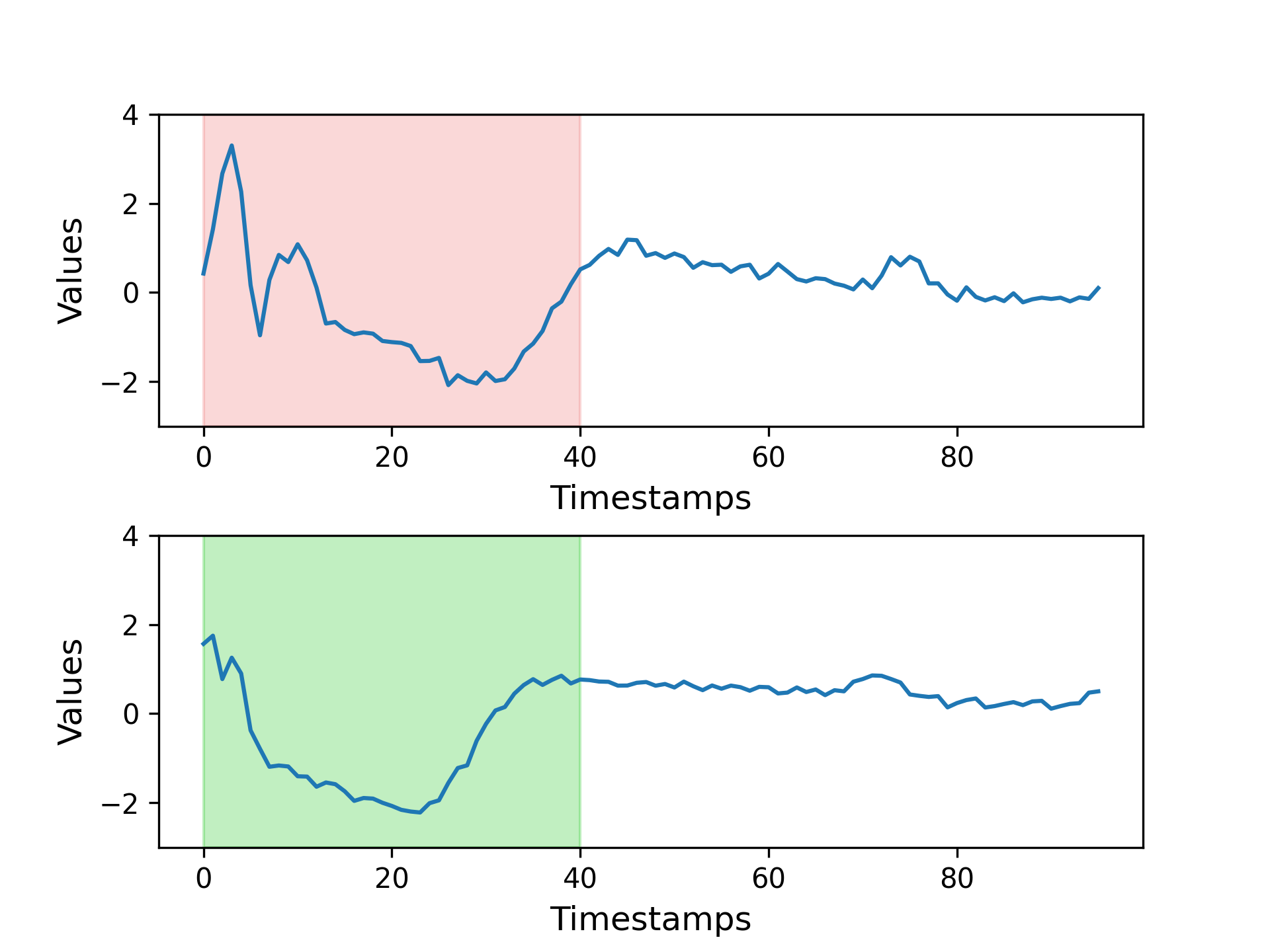}
\vspace{-2mm}
\caption{Example of Shapelets that distinguish two classes}
\vspace{-4mm}
\label{Shapelets}
\end{figure}

Time series shapelets are local discriminative subsequences that are accurate and interpretable for the classification problem of time series~\cite{li2021shapenet}. In Figure~\ref{Shapelets}, a visualization of two classes of time series is provided. It should be noted that the main divergence between these two is predominantly manifested in the highlighted area, \ie the time series shapelet. Typically, TSC models can directly apply the top $k$ shapelets as features. In our case, we conduct a class-wise adversarial attack using the top one shapelet interval of each time series class, assuring a more niche-targeting attack.

Shapelets $S$ are subsequences of the contained time series in $D$. A shapelet $s$ in $S$ tries to divide the dataset $D$ into two disjoint datasets $D_1$ and $D_2$ with a large information gain~\cite{ye2009time}. Specifically, $D_1=\{x: x\in{D}, subdist(s, x)\leq{\tau}\}$ and $D_2=\{x: x\in{D}, subdist(s, x)>{\tau}\}$, where $subdist$ represents the minimum distance between shapelet $s$ and the subsequences in time series $x$, and $\tau$ is the distance threshold. Assuming that there are $T$ time series in $D$ with $|C|$ classes and each class has $t_i$ instances in $D$. The information gain (IG) is as follows:

\vspace{-4mm}
\begin{align}
    E(D) = -\sum_{i=1}^{|C|}{\frac{t_i}{T}\log{\frac{t_i}{T}}}
\end{align}
\vspace{-2mm}

\vspace{-4mm}
\begin{align}
    I_s = E(D)-\frac{|D_1|}{|D|}E(D_1)-\frac{|D_2|}{|D|}E(D_2)
    \label{Information Gain}
\end{align}
\vspace{-2mm}

where $E(D)$ is the entropy and the information gain can be computed as the entropy increase after the division. Intuitively, a shapelet is a subsequence that can best distinguish different categories. Thus, shapelets may provide a more effective and imperceptible perturbed interval than randomly selected subsequences. In our case, we utilize the best shapelets of each class to determine the main perturbed intervals.

\textbf{Theorem 1.} \emph{Suppose we denote the gradient of the time series $x$ with respect to the loss function $L(\cdot)$ as ${\nabla}L(x)$. If we conduct perturbations on the time series shapelet, to achieve the optimal performance of increasing the classification loss function, the perturbation can be computed as follows:}

\begin{figure}
\centering
\includegraphics[width=.9\linewidth]{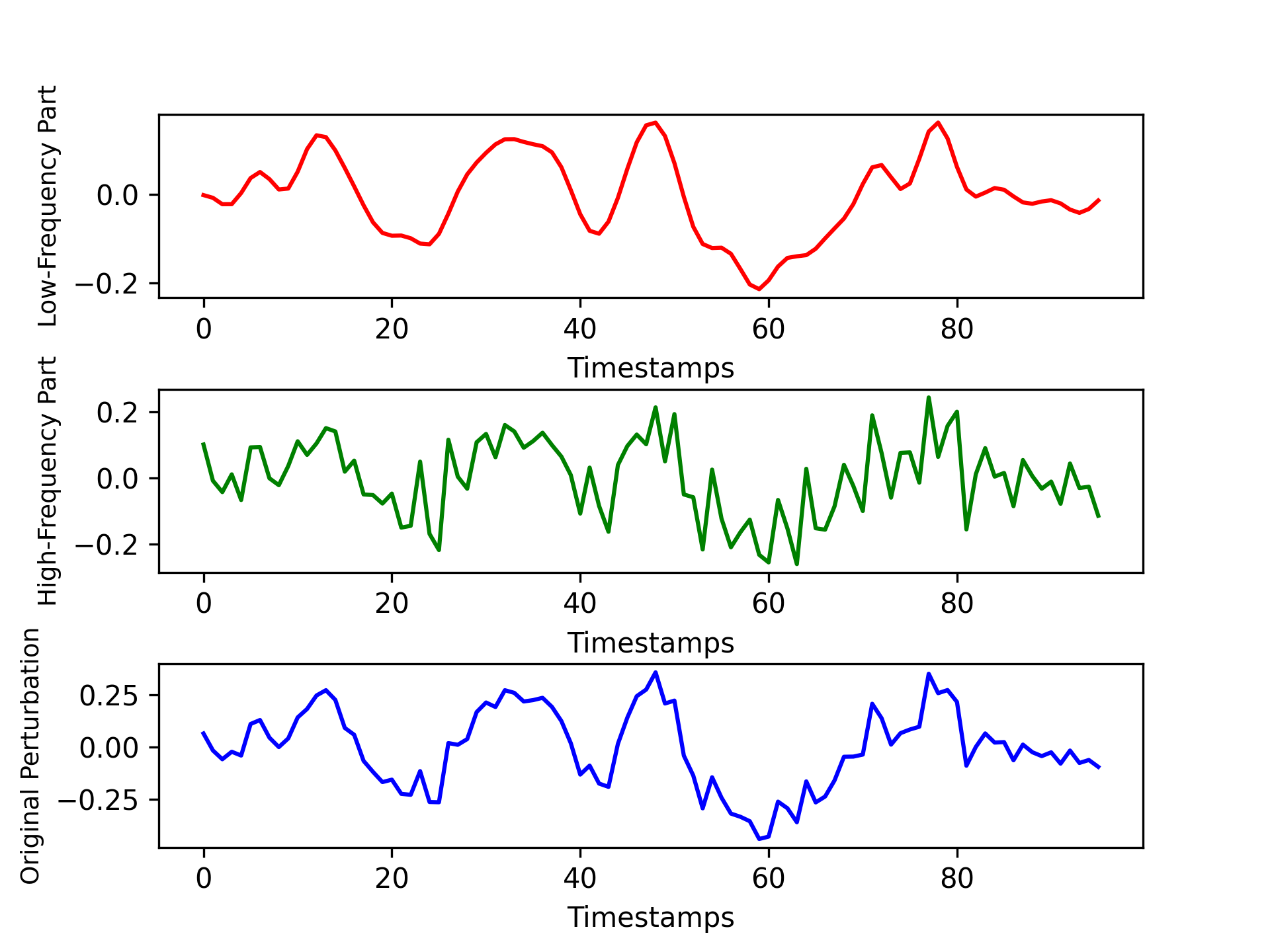}
\vspace{-2mm}
\caption{Frequency transform that decomposes perturbation into two parts, namely, low and high-frequency components}
\vspace{-2mm}
\label{Frequency}
\end{figure}

\vspace{-4mm}
\begin{align}
    x^{adv} = x+{\epsilon}\cdot{\nabla}L_{n}(s)
\end{align}
\vspace{-2mm}

\vspace{-4mm}
\begin{align}
    {\nabla}L(x) = {\nabla}L(s)+{\nabla}L(\hat{x})
    \label{Decompose}
\end{align}
\vspace{-2mm}

\noindent \emph{where ${\nabla}L(s)$ and ${\nabla}L(\hat{x})$ denotes the gradients in the interval of shapelet $s$ and the remaining areas, respectively. While ${\nabla}L_{n}(s)$ stands for the normalized ${\nabla}L(s)$.}

According to the theorem, the best performance is achieved when the perturbation on the shapelet interval is in the same direction as the shapelet part of the original gradient. The proof is given in the appendix. 

We focus more on the discriminative segments of the time series that make the attack effective with local perturbations. However, to avoid the TSC model retaining memory on the other segments except for shapelet parts, a very small and nearly ignorable perturbation should also be conducted across the non-shapelet segments, \ie the perturbation can be set to $\lambda$ (in our case, 0.1) times of the shapelet part, ensuring that the perturbation remains minor and imperceptible. The perturbation $p$ can be computed as follows:

\vspace{-4mm}
\begin{align}
    p=\frac{{\nabla}L(s)}{||{\nabla}L(s)||_2}+\lambda\cdot\frac{{\nabla}L(\hat{x})}{||{\nabla}L(\hat{x})||_2}
    \label{Perturbation}
\end{align}
\vspace{-2mm}

\textbf{Theorem 2.} \emph{With the same constraint of $L_2$ norm that $||G_{s}(x)-x||_2\leq\epsilon$ and $||G_{n}(x)-x||_2\leq\epsilon$ and denote the loss function and its increment as $L(\cdot)$ and ${\Delta}L(\cdot)$, respectively. Then, $G_{s}(x)$ is more effective than $G_{n}(x)$:}

\vspace{-4mm}
\begin{align}
{\Delta}L(G_{s}(x))>{\Delta}L(G_{n}(x))
\end{align}
\vspace{-4mm}

\noindent \emph{where $G_{s}(x)$ and $G_{n}(x)$ denotes the attacks on the shapelet and the non-shapelet part of the time series.} 

This theorem supports the idea that adversarial attacks on time series shapelets can enhance performance and improve imperceptibility. The detailed proof is also given in the appendix.

\subsubsection{Frequency-Domain Attack}

Intuitively, the high-frequency components representing noise manifested as oscillation, are more imperceptible than the low-frequency components, \eg direct current (DC) components demonstrated as a level shift in time series. Thus, even though local perturbations enhance imperceptibility, there is still a risk that these perturbations contain low-frequency components, \eg square wave signals, which may be perceptible to the human visual system (HVS)~\cite{yi2024time, luo2022frequency}. Therefore, a constraint that limits these local perturbations into imperceptible high-frequency areas is needed.

We adopt the widely used time-frequency analysis tool discrete cosine transform (DCT). An example is shown in Figure~\ref{Frequency}, where the original perturbation is decomposed into easily perceptible low-frequency components and more imperceptible high-frequency counterparts. Specifically, DCT transforms the perturbation series $p$ into a length $T$ frequency spectrum, which can be formulated as follows:

\vspace{-4mm}
\begin{align}
    w_k = d_k\sum_{n=0}^{T-1}{p_n\cdot\cos(\frac{\pi(2n+1)k}{2T})}
    \label{DCT}
\end{align}
\vspace{-2mm}

\vspace{-4mm}
\begin{align}
    d_{k}=\left\{
    \begin{aligned}
    \frac{1}{\sqrt{T}} & , & k=0, \\
    \frac{2}{\sqrt{T}} & , & k>0.
\end{aligned}
\right.
\end{align}
\vspace{-2mm}

\noindent where $w_k$ is the $k^{th}$ components of the frequency spectrum. Then, we filter out the low-frequency components, \ie setting the DCT coefficients corresponding to the low frequencies to zero. Finally, we compute the perturbation $p^{adv}$ after high-pass filtering by applying the inverse DCT:

\vspace{-4mm}
\begin{align}
    p^{adv}_n = \sum_{k=\gamma}^{T-1}{d_k{\cdot}w_k\cdot\cos(\frac{\pi(2n+1)k}{2T})}
    \label{IDCT}
\end{align}
\vspace{-2mm}

\noindent where $p^{adv}_n$ represents the $n^{th}$ values of the perturbation and $\gamma$ represents the cutoff frequency. It should be noted that to ensure $L_2$ remains unchanged, we also apply a scaling on the filtered perturbation.

\subsubsection{SFAttack}

In SFAttack, we adopt a novel similarity-based loss function. Instead of maximizing the overall cross-entropy loss, we focus on the class that is most similar to the true class. Besides, different from existing approaches~\cite{ismail2019deep}, we utilize the $L_2$ norm constraint and FGM-based optimizer for better effectiveness and imperceptibility. 

Specifically, conventional adversarial attack methods solve the problem of maximizing the classification loss within a certain constraint, which can be written as follows:

\vspace{-4mm}
\begin{align}
    x^{adv}=\mathop{\arg\max}\limits_{||x^{adv}-x||_2\leq{\epsilon}}{L(x)}
\end{align}
\vspace{-2mm}
 
In other words, this encourages adversarial examples to maximize classification loss, and thus, they are misclassified as classes other than the true class $y$. However, this potentially weakens the effectiveness of the attack as it results in dispersion in imitating all other categories rather than focusing on the easiest misclassified class. Our solution is straightforward targeting the most similar 'other class' because it is easier to make the classifier confuse two similar categories due to their inherent similarities. Thus, the required adversarial perturbation may be smaller and harder for defensive mechanisms to detect. This can be written as follows: 

\vspace{-4mm}
\begin{align}
    x^{adv}_j=\mathop{\arg\max}\limits_{||x^{adv}-x||_2\leq{\epsilon}}{f_k(x)}
\end{align}
\vspace{-1mm}

\vspace{-4mm}
\begin{align}
    k=\mathop{\arg\min}\limits_{y_j{\neq}y_k}{\frac{x_j{\cdot}x_k}{||x_j||_2{\cdot}||x_k||_2}}
\end{align}
\vspace{-1mm}

\noindent where $x^{adv}_j$ represents the adversarial example for the $j^{th}$ sample $x_j$. The index of the most similar sample from the other class from $x_j$ is denoted as $k$ and $f_k(x)$ stands for the $k^{th}$ component of the output of classifier $f(\cdot)$. 

We adopt the $L_2$ norm constraint and FGM-based optimizer for generating adversarial examples in SFAttack. The following two theorem shows the motivation of our choice of constraint and optimizer:

\textbf{Theorem 3.} \emph{Denote the $G_{FGM}(x)$ and $G_{FGSM}(x)$ as the adversarial example generated by FGM and FGSM, respectively. Given the same constraint of $L_2$ norm that $||G_{FGM}(x)-x||_2\leq\epsilon$ and $||G_{FGSM}(x)-x||_2\leq\epsilon$ and denotes the low-frequency component of a time series as $LF(\cdot)$, $G_{FGM}(x)$ contains less low-frequency component than $G_{FGSM}(x)$:} 

\vspace{-4mm}
\begin{align}
    LF(G_{FGM}(x))<LF(G_{FGSM}(x))
\end{align}
\vspace{-6mm}

Two adversarial samples generated by both FGSM and our SFAttack based on FGM in the EGG dataset are shown in Figure~\ref{Examples}. We can observe that the perturbations introduced by FGSM predominantly contain low-frequency components, such as square waves, which are more perceptible to the human eye, thereby reducing the imperceptibility of the attack. On the contrary, our SFAttack maintains tiny perturbations across the majority of the time series, while in some representative parts of the time series, our method generates a more oscillatory signal with varying amplitude instead of a consistent value compared to FGSM. The proof is given in the appendix. 

\textbf{Theorem 4.} \emph{Given the same constraint of $L_2$ norm that $||G_{FGM}(x)-x||_2\leq\epsilon$ and $||G_{FGSM}(x)-x||_2\leq\epsilon$ and denote the loss function of the TSC classifier $f$ and its increment as $L(\cdot)$ and ${\Delta}L(\cdot)$, respectively. Then, $G_{FGM}(x)$ is more effective than $G_{FGSM}(x)$:}

\vspace{-4mm}
\begin{align}
    {\Delta}L(G_{FGM}(x))>{\Delta}L(G_{FGSM}(x))
\end{align}
\vspace{-6mm}

This theorem further supports our motivation of applying $L_2$ norm constraints in generating adversarial samples. That is to say, using $L_2$ norm as the constraint can not only achieve higher imperceptibility as the perturbations include less easily perceived low-frequency components but is also more effective in increasing the loss function of the target classifier. The proof is also given in the appendix.

Our overall attacking algorithm is shown in Algorithm~\ref{Algo1}. We first compute the class-wise shapelet and obtain the similarity-based loss function. The gradient to the loss will be decomposed into the shapelet part and non-shapelet part, then the perturbation can be computed as Equation~\eqref{Perturbation}. Then DCT will be conducted as shown in Equation~\eqref{DCT} and~\eqref{IDCT} to obtain the final perturbation and adversarial examples.  

\begin{algorithm}[t]
    \caption{Shapelet-based Frequency-domain Attack}
    \label{Algo1}
    \normalsize
    \begin{algorithmic}[1] 
        \REQUIRE {Input time series dataset $X$ and its ground-truth $y$} 
        \REQUIRE {The classifier $f$, attack budget $\epsilon$ and iteration $I$}
        \REQUIRE {The loss function $L$, the hyperparameter $\gamma$ and $\lambda$}
        \ENSURE {The adversarial time series $X_{adv}$}
        \STATE {$\alpha=\frac{\epsilon}{T}$, $X_0=X$}
        \STATE {Compute the shapelets $S$ for each class $\triangleright$ Eq.~\eqref{Information Gain}}
        \FOR {$n = 1 $; $ n \leq N $; $ n ++ $}
        \STATE {$l=\mathop{\arg\min}\limits_{y_j{\neq}y_l}{\frac{x_j{\cdot}x_l}{||x_j||_2{\cdot}||x_l||_2}}$}
        \STATE {$L=1-f_l(x)$}
        \FOR {$i = 0 $; $ i \leq I $; $ i ++ $}
        \STATE {Compute the gradient ${\nabla}L(x^i_n)$ to $L(x^i_n)$}
        \STATE {${\nabla}L(x^i_n) = {\nabla}L(s^i_n)+{\nabla}L(\hat{x}^i_n)$ $\triangleright$ Eq.~\eqref{Decompose}}
        \STATE {$p^i_n=\frac{{\nabla}L(s^i_n)}{||{\nabla}L(s^i_n)||_2}+\lambda\cdot\frac{{\nabla}L(\hat{x}_n^i)}{||{\nabla}L(\hat{x}_n^i)||_2}$ $\triangleright$ Eq.~\eqref{Perturbation}}
        \STATE {$w^i_k = d^i_k\sum_{t=0}^{T-1}{{p^i_n}(t)\cdot\cos(\frac{\pi(2n+1)k}{2T})}$ $\triangleright$ Eq.~\eqref{DCT}}
        \STATE {$p^{i}_n = \sum_{k=\gamma}^{T-1}{d^i_k{\cdot}w^i_k\cdot\cos(\frac{\pi(2n+1)k}{2T})}$ $\triangleright$ Eq.~\eqref{IDCT}}
        \STATE {$x^{i+1}_n=x^i_n+\alpha\cdot{p^i_n}$}
        \ENDFOR
        \ENDFOR
        \STATE {$X_I=\left[x^I_1, x^I_2,...,x^I_N\right]$}
        \RETURN {$X_I$}
    \end{algorithmic} 
\end{algorithm}
\setlength\textfloatsep{1mm}

%% file: contents/04_experiments.tex
\section{Experiments}

In this section, we conduct extensive experiments to validate the effectiveness of our proposed Shapelet-based Frequency-domain Attack (SFAttack). We clarify the setup of the experiments first. Then, we demonstrate the white-box attacking performance and the imperceptibility measures of our method against competitive baseline methods. The experiment results demonstrate that our method largely improves the imperceptibility of adversarial examples while maintaining an attack performance that is on par with established baseline methods. Furthermore, we present the ablation study on the shapelet and frequency transform components to study their impact on the attacking performance and imperceptibility. We have also conducted a parameter analysis of the hyperparameters used in SFAttack. 

\subsection{Experiment Setup}

We illustrate our experimental setting on dataset selection, target TSC models, baseline methods, evaluation metrics, and our implementation in this section.

\textbf{Datasets.} We conduct the experiments on three datasets: ECG, Uwave, and Phoneme. These datasets are from the most widely used TSC benchmark UCR~\cite{dau2019ucr} that represent three distinct application areas. Specifically, the ECG dataset is used to analyze heartbeats and detect anomalies in cardiac patterns. The Uwave dataset is utilized in gesture recognition tasks. The Phoneme dataset involves the classification of phonetic sounds for speech recognition.

\textbf{Targets.} We choose three representative TSC models containing TCN~\cite{bai2018empirical}, InceptionTime~\cite{ismail2019deep}, and MiniRocket~\cite{tan2022multirocket} as the target model to construct adversarial time series and test these models under the white-box setting. Specifically, TCN and InceptionTime are the two most widely-used deep learning methods while the MiniRocket falls in the category of machine learning. 

\begin{table*}[htbp]
\centering
\scalebox{.9}{\begin{tabular}{cccccccccc}
\toprule
Dataset & Model & Metrics & FGSM & BIM & FGM & PGD & C\&W & SGM & SFAttack\\ 
\midrule

\multirow{12}{*}{ECG} & 
\multirow{4}{*}{TCN} & ASR(\%) $\uparrow$ & 76.23 & 78.30 & 82.14 & \textbf{84.63} & 81.76 & 80.42 & 83.47\\
& & $L_2$$ \downarrow$ & 15.82 & 14.90 & 13.43 & 13.28 & 15.67 & 14.25 & \textbf{10.34}\\
& & DTW $\downarrow$ & 215.74 & 244.18 & 200.81 & 194.32 & 201.47 & 209.61 & \textbf{150.50}\\
& & LF $\downarrow$ & 4.86 & 4.58 & 4.31 & 4.10 & 5.49 & 5.02 & \textbf{2.27}\\
\cmidrule{2-10}
& \multirow{4}{*}{InceptionTime} & ASR(\%) $\uparrow$ & 78.31 & 78.69 & 81.43 & \textbf{84.09} & 81.32 & 82.09 & 82.60\\
& & $L_2$ $\downarrow$ & 17.91 & 17.04 & 15.49 & 14.78 & 16.62 & 15.77 & \textbf{13.25}\\
& & DTW $\downarrow$ & 204.89 & 217.64 & 220.76 & 217.45 & 197.85 & 209.43 & \textbf{167.42}\\
& & LF $\downarrow$ & 5.17 & 5.48 & 5.67 & 5.81 & 5.43 & 5.17 & \textbf{2.41}\\
\cmidrule{2-10}
& \multirow{4}{*}{MiniRocket} & ASR(\%) $\uparrow$ & 77.45 & 78.91 & 78.42 & 80.42 & 81.97 & 82.45 & \textbf{84.57}\\
& & $L_2$ $\downarrow$ & 17.85 & 18.69 & 19.42 & 17.34 & 16.54 & 17.08 & \textbf{14.21}\\
& & DTW $\downarrow$ & 214.78 & 220.10 & 216.74 & 232.13 & 206.42 & 197.91 & \textbf{162.42}\\
& & LF $\downarrow$ & 5.14 & 4.76 & 5.42 & 4.67 & 5.62 & 4.99 & \textbf{2.12}\\

\midrule

\multirow{12}{*}{UWave} & 
\multirow{4}{*}{TCN} & ASR(\%) $\uparrow$ & 71.45 & 73.51 & 74.37 & 70.34 & 69.06 & 73.26 & \textbf{75.61}\\
& & $L_2$ $\downarrow$ & 18.57 & 17.64 & 19.32 & 18.05 & 20.39 & 22.15 & \textbf{14.40}\\
& & DTW $\downarrow$ & 199.78 & 201.45 & 205.74 & 203.34 & 237.1 & 224.8 & 140.77\\
& & LF $\downarrow$ & 5.65 & 5.34 & 5.10 & 4.89 & 5.26 & 5.95 & \textbf{2.79}\\
\cmidrule{2-10}
& \multirow{4}{*}{InceptionTime} & ASR(\%) $\uparrow$ & 72.46 & 74.91 & 75.62 & \textbf{76.23} & 74.73 & 75.61 & 75.44\\
& & $L_2$ $\downarrow$ & 20.78 & 19.75 & 18.46 & 17.75 & 21.63 & 22.97 & \textbf{13.61}\\
& & DTW $\downarrow$ & 153.20 & 165.56 & 142.92 & 152.27 & 153.03 & 158.26 & \textbf{125.92}\\
& & LF $\downarrow$ & 5.13 & 5.75 & 5.39 & 6.05 & 5.67 & 5.53 & \textbf{2.86}\\
\cmidrule{2-10}
& \multirow{4}{*}{MiniRocket} & ASR(\%) $\uparrow$ 
& 72.37 & 71.08 & 70.11 & 72.56 & 71.45 & 73.63 & \textbf{74.89}\\
& & $L_2$ $\downarrow$ & 22.71 & 18.91 & 19.74 & 23.76 & 20.78 & 24.12 & \textbf{15.31}\\
& & DTW $\downarrow$ & 149.36 & 156.82 & 139.56 & 142.78 & 157.89 & 146.98 & \textbf{114.71}\\
& & LF $\downarrow$ & 5.52 & 5.91 & 6.48 & 6.13 & 5.41 & 5.27 & \textbf{3.09}\\

\midrule

\multirow{12}{*}{Phoneme} & 
\multirow{4}{*}{TCN} & ASR(\%) $\uparrow$ & 82.23 & 81.30 & 83.44 & 84.97 & 86.06 & 87.43 & \textbf{90.13}\\
& & $L_2$$ \downarrow$ & 5.05 & 6.03 & 5.92 & 5.79 & 5.95 & 5.91 & \textbf{4.76}\\
& & DTW $\downarrow$ & 25.74 & 27.18 & 24.51 & 24.32 & 23.73 & 27.68 & \textbf{20.59}\\
& & LF $\downarrow$ & 1.08 & 1.11 & 1.29 & 1.26 & 1.43 & 1.20 & \textbf{0.69}\\
\cmidrule{2-10}
& \multirow{4}{*}{InceptionTime} & ASR(\%) $\uparrow$ & 84.62 & 83.09 & 86.25 & 87.13 & 84.56 & 84.89 & \textbf{89.54}\\
& & $L_2$ $\downarrow$ & 5.31 & 6.04 & 5.96 & 5.70 & 6.01 & 5.84 & \textbf{4.74}\\
& & DTW $\downarrow$ & 25.41 & 26.78 & 24.73 & 24.39 & 28.26 & 25.43 & \textbf{19.43}\\
& & LF $\downarrow$ & 1.38 & 1.61 & 1.25 & 1.32 & 1.12 & 1.44 & \textbf{0.62}\\
\cmidrule{2-10}
& \multirow{4}{*}{MiniRocket} & ASR(\%) $\uparrow$ & 82.43 & 84.31 & 88.75 & 91.47 & 91.04 & 89.32 & \textbf{91.69}\\
& & $L_2$ $\downarrow$ & 5.97 & 5.89 & 5.84 & 6.25 & 6.13 & 5.84 & \textbf{4.59}\\
& & DTW $\downarrow$ & 29.75 & 26.49 & 24.37 & 25.49 & 27.54 & 27.89 & \textbf{21.71}\\
& & LF $\downarrow$ & 1.64 & 1.78 & 1.87 & 1.61 & 1.52 & 1.68 & \textbf{0.71}\\
\midrule

\end{tabular}}
\vspace{-0mm}
\caption{The white-box attack performance on three datasets. ASR is the attack success rate (\%) and $L_2$ represents the $L_2$ norm errors. DTW is the dynamic time-warping distance that measures the similarity between two temporal sequences that may vary in speed or out of sync. LF is the $l_2$ norm of the low-frequency component. The best result is in bold.}
\vspace{-6mm}
\label{Performance}
\end{table*}

\textbf{Baselines.} We compare our approach with six baselines: FGSM, BIM~\cite{ismail2019deep}, FGM, PGD~\cite{madry2018towards}, C\&W~\cite{carlini2017towards} and SGM~\cite{pialla2022smooth}. Among them, SGM is specially designed for TSC models by incorporating a fused LASSO term for smoother perturbations. We compare our approach with these baselines on both effectiveness and imperceptibility to validate the effectiveness of our method.

\textbf{Metrics.} Besides the evaluation metric on effectiveness, namely ASR (Attack Success rate), we evaluate the imperceptibility of the crafted adversarial time series from three perspectives, namely, the $l_2$ distance, dynamic time warping (DTW) distance, and low-frequency component (LF). The $l_2$ distance is straightforward for evaluating imperceptibility between the original and adversarial time series. DTW distance aligns time series in a non-linear manner, making it more robust for comparing the shapes and patterns of the original and adversarial time series. In contrast, the LF metric evaluates the ratio of HVS perceptible low-frequency components.

\textbf{Implementation.} For a fair comparison, we set the parameter of attack budget $\epsilon=0.25$ for all the attacking methods. Besides, the number of iterations is set to be $I=10$, and the shapelet length ratio $R=0.5$. The coefficient $\lambda$ and cutoff frequency are 0.1 and $\frac{T}{4}$. We adopt the untargeted attack strategy to evaluate the imperceptibility and attacking performance. We follow the hyper-parameter settings of all the baselines in their publicly available implementations. All the experiments are conducted on a Linux server with Intel Xeon Gold 6140 CPU @ 2.30GHZ and Tesla V100 PCIe GPU. 

\subsection{The Effectiveness of SFAttack}

In this section, we analyze the performance of our proposed approach SFAttack against the state-of-the-art baselines from the perspective of imperceptibility and attack success rate, respectively. Specifically, we attack a given model and directly test the model by crafting adversarial samples, also known as the white-box setting.

As shown in Table~\ref{Performance}, our approach achieves the highest performance on most of the target models, achieving the best average white-box ASR of 83.7\%  compared with all the baselines. Furthermore, SFAttack outperforms all the other baselines on all three measurements of imperceptibility, demonstrating the high quality of adversarial time series generated by our approach. In particular, we can achieve comparable performance with state-of-the-art baselines while reducing the perturbation magnitude ($l_2$ loss) at about 17.4\%. An interesting observation is that the performance improvement on UWave and Phoneme datasets is greater than on ECG. This is because, due to our design of the similarity-based loss function, when the number of classes is larger than two, our method can focus more on the easily misclassified classes. However, the ECG dataset contains only two classes, limiting the potential for such improvements. 

From the perspective of imperceptibility, utilizing shapelets that focus on the discriminative segments of the time series proves to be highly effective. It not only helps maintain high performance but also significantly reduces the $L_2$ norm of the perturbations, thereby enhancing the imperceptibility of the adversarial time series. Moreover, the application of the DCT provides further improvements in terms of imperceptibility by reducing the HVS-perceptible LF components. Though SGM~\cite{pialla2022smooth} considers taking the fused lasso term into the loss function to provide a smoother perturbation, it overlooks the local features and frequency-domain information of time series, which hinders it from generating high-quality adversarial examples.  

\begin{figure}
  \centering
  \subfigure[Attack Success Rate]{
  \label{ASR_I}
  \includegraphics[width=0.42\columnwidth]{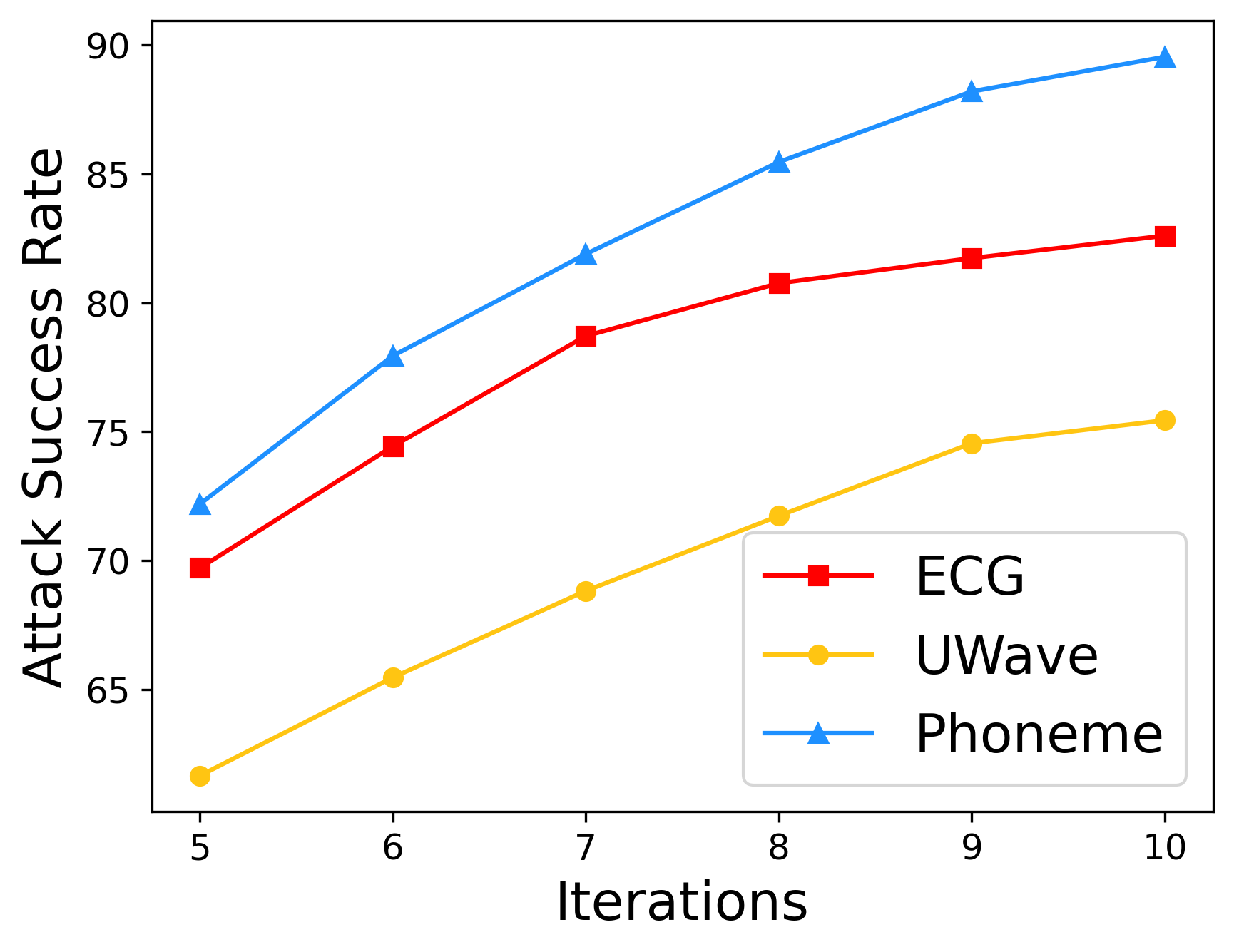}
  }
  \subfigure[$L_2$ Norm Distance]{
  \label{L2_I}
  \includegraphics[width=0.42\columnwidth]{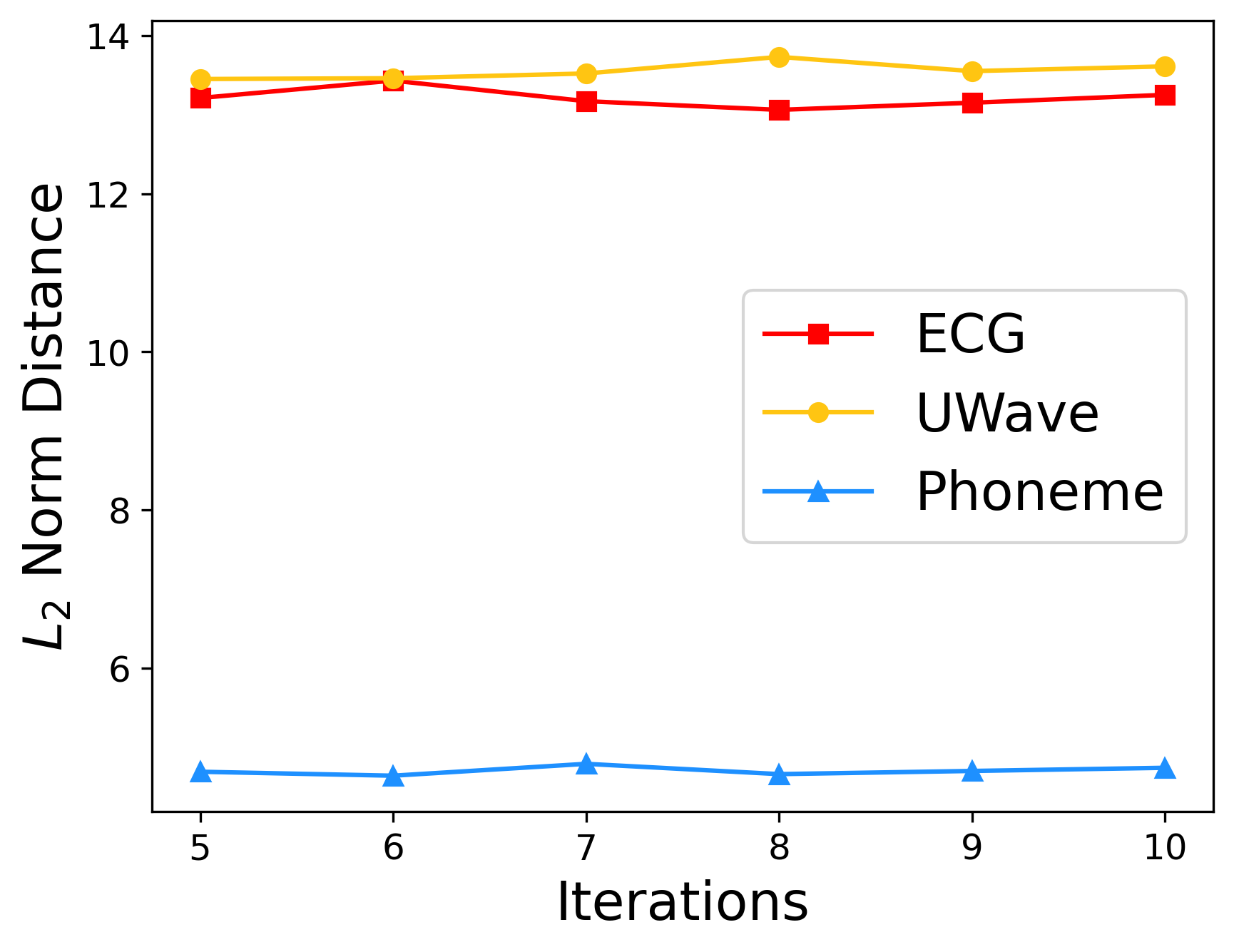}
  }
  \subfigure[DTW Distance]{
  \label{DTW_I}
  \includegraphics[width=0.42\columnwidth]{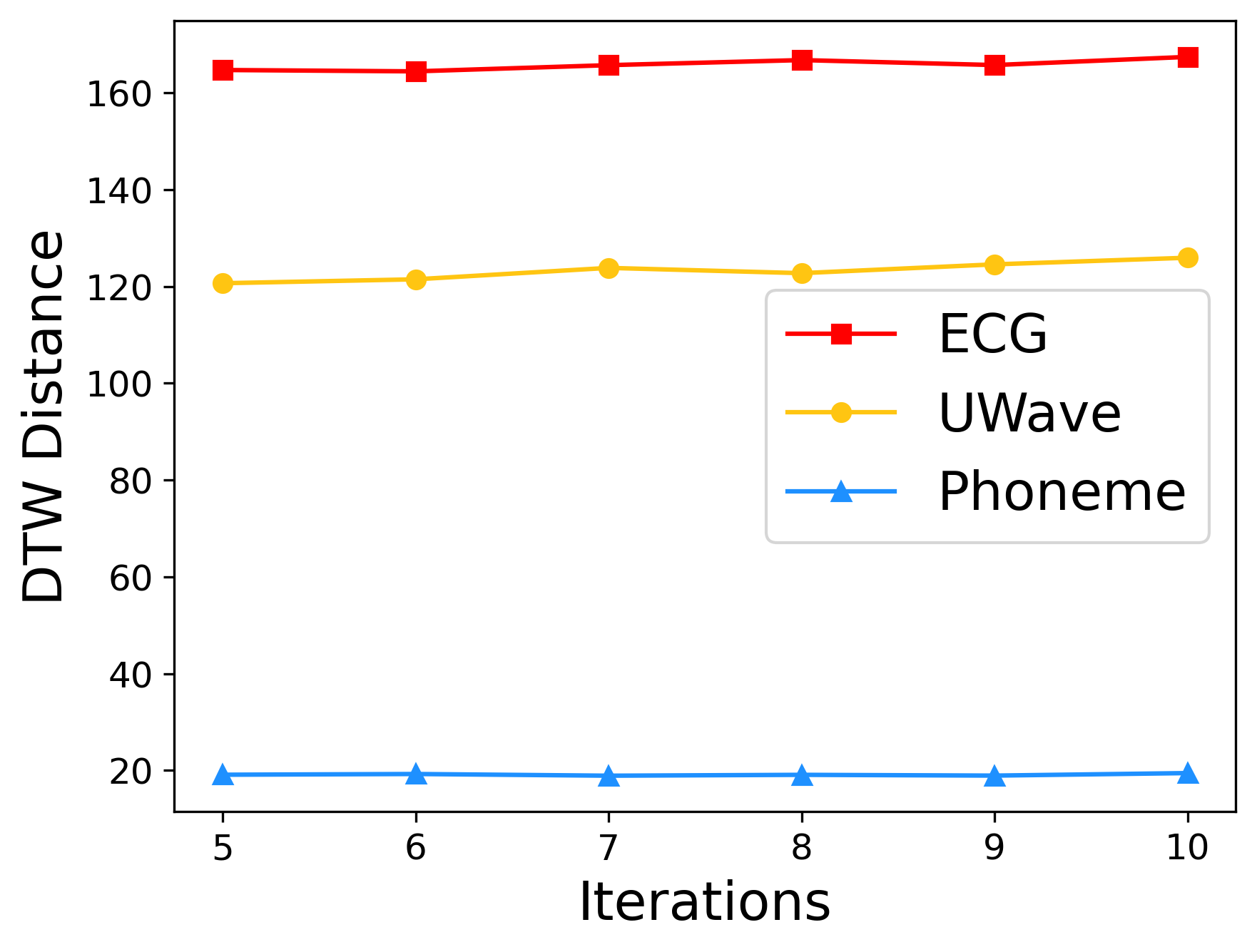}
  }
  \subfigure[LF Components]{
  \label{LF_I}
  \includegraphics[width=0.42\columnwidth]{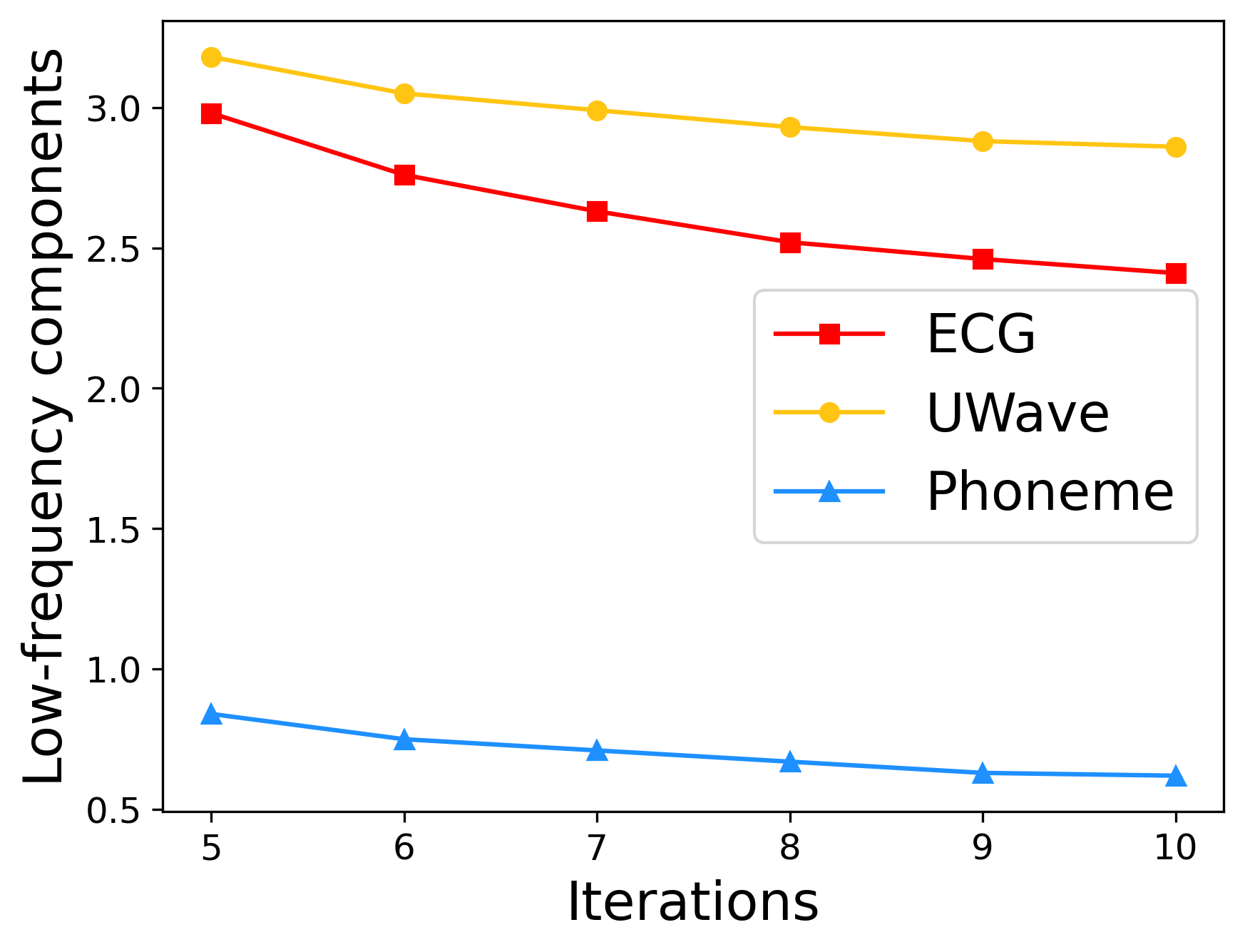}
  }
  \vspace{-0mm}
  \caption{Ablation study on iteration on InceptionTime model}
  \vspace{-0mm}
  \label{Iteration}
\end{figure}

\begin{table}[t]
\centering
\scalebox{.9}{\begin{tabular}{cccccc}
\toprule
Dataset & Metrics & $\textit{w/o}$ S & $\textit{w/o}$ D & $\textit{w/o}$ L & SFAttack\\ 
\midrule
 
\multirow{4}{*}{ECG} & ASR & 81.93 & 82.46 & 80.18 & 82.60\\
& $L_2$ & 15.28 & 14.67 & 13.40 & 13.25\\
& DTW & 204.32 & 186.47 & 169.89 & 167.42\\
& LF & 3.10 & 5.39 & 2.42 & 2.41\\
\bottomrule

\end{tabular}}
\vspace{2mm}
\caption{Ablation study with three variant baselines}
\vspace{-0mm}
\label{Ablation}
\end{table}

\subsection{Ablation Study}

We derive the first variant baseline (\textit{w/o} S) by replacing the shapelets with randomly selected segments of the same length. The second and third variants remove the DCT (\textit{w/o} D) and similarity-based loss function (\textit{w/o} L) from SFAttack, respectively. The performance comparison between SFAttack and its variants is shown in Table~\ref{Ablation}. On one hand, we can observe that the ASR of SFAttack is enhanced by the similarity-based loss function in all datasets. On the other hand, the shapelet extraction and DCT contribute to the imperceptibility of the adversarial time series in two different ways. Specifically, the use of shapelets contributes to reducing the magnitude of the perturbations ($L_2$ norm) by focusing on the discriminative local patterns of the time series. While the application of DCT helps to reduce the low-frequency components within the perturbations. These two strategies complement each other to ensure that the generated adversarial samples are both effective in fooling the classifier and imperceptible to HVS.

\begin{figure}
  \centering
  \subfigure[Attack Success Rate]{
  \label{ASR_B}
  \includegraphics[width=0.42\columnwidth]{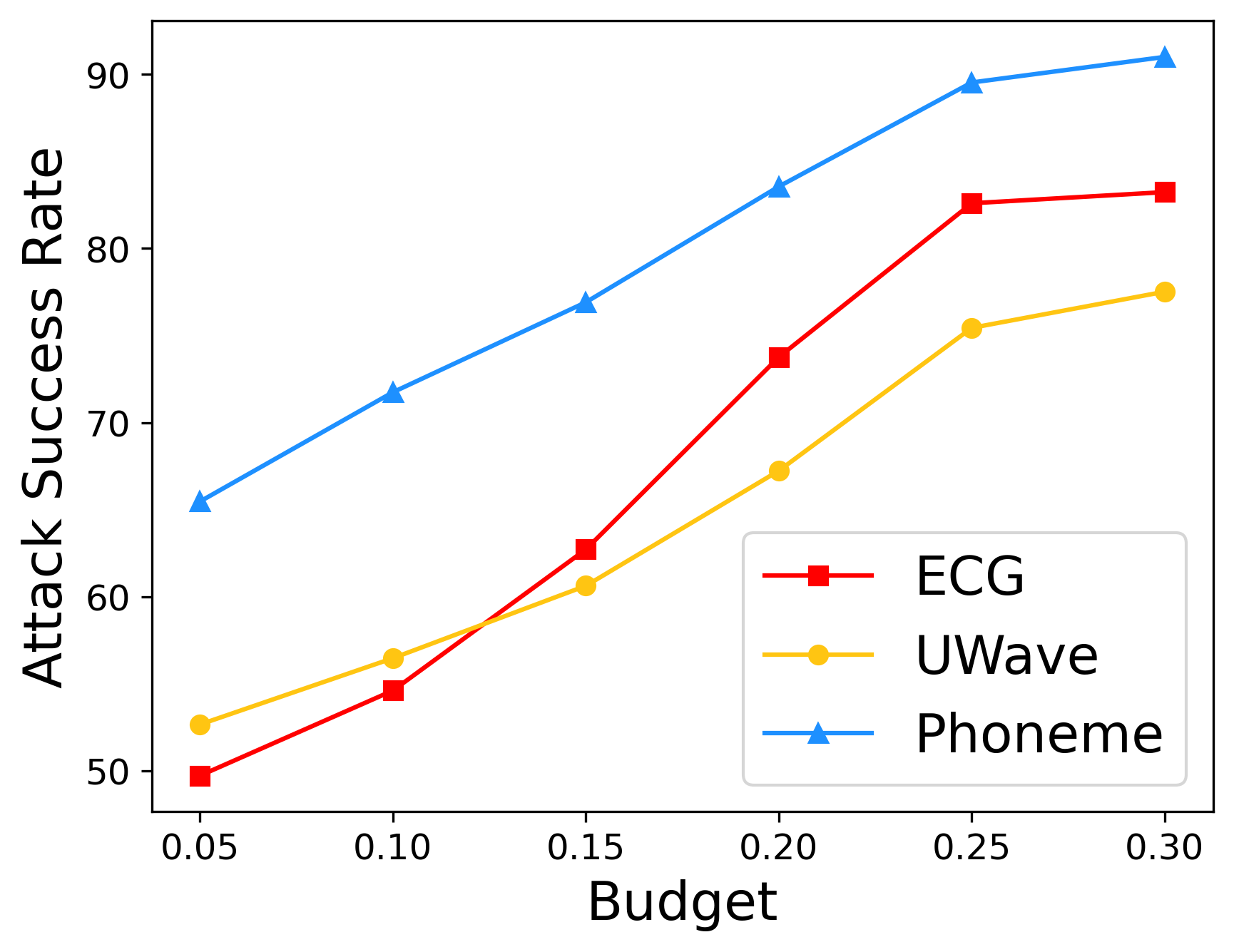}
  }
  \subfigure[$L_2$ Norm Distance]{
  \label{L2_B}
  \includegraphics[width=0.42\columnwidth]{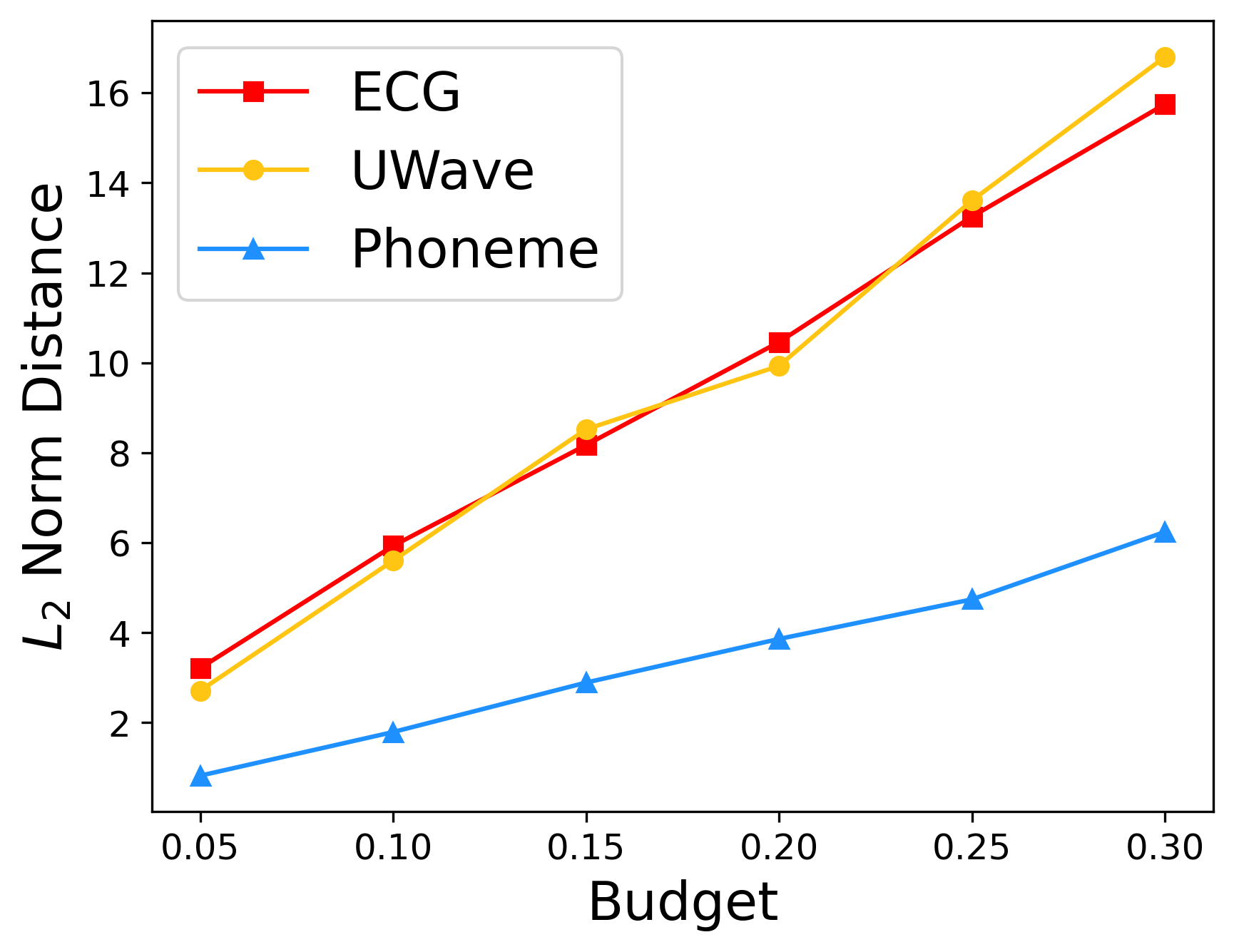}
  }
  \subfigure[DTW Distance]{
  \label{DTW_B}
  \includegraphics[width=0.42\columnwidth]{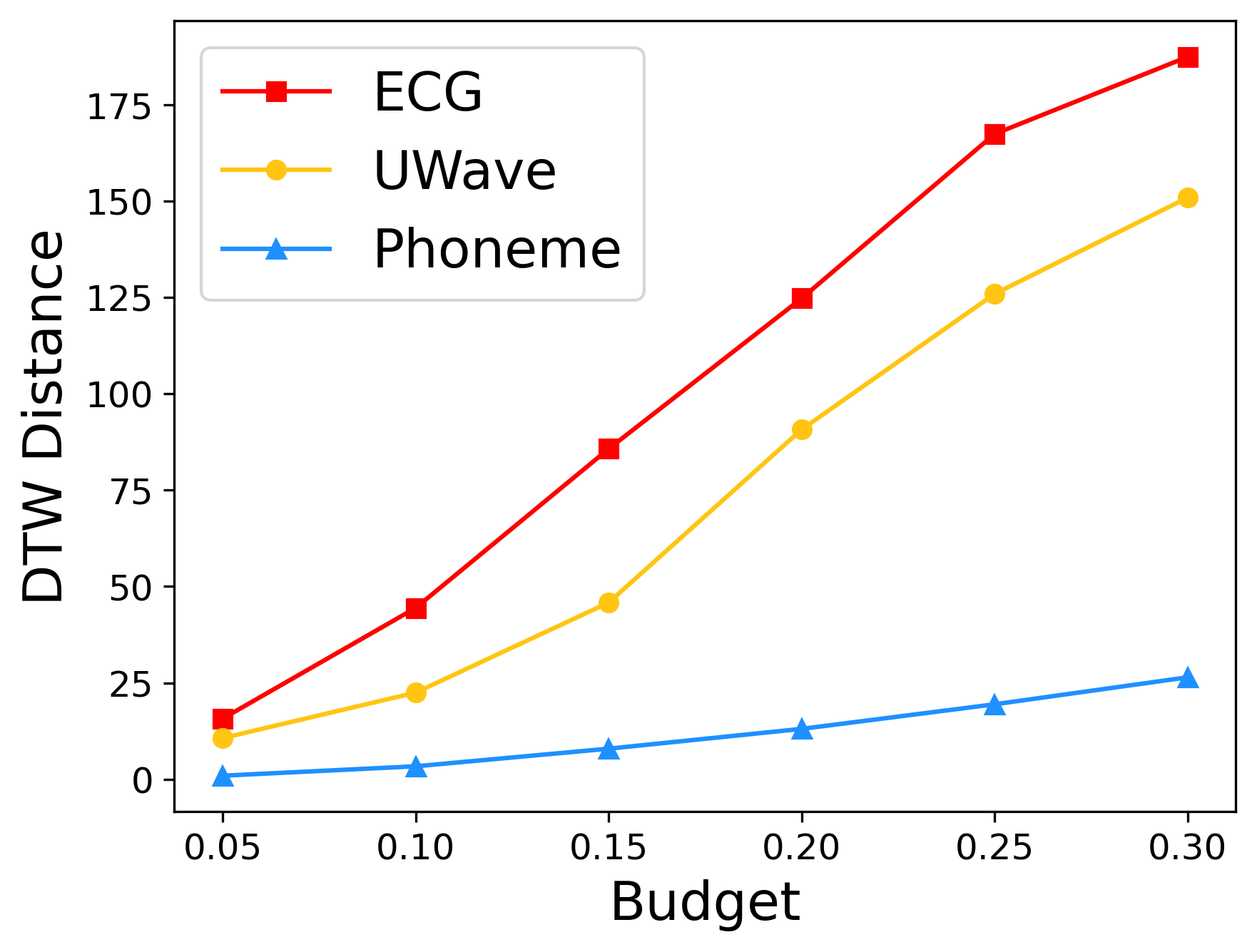}
  }
  \subfigure[LF Components]{
  \label{LF_B}
  \includegraphics[width=0.42\columnwidth]{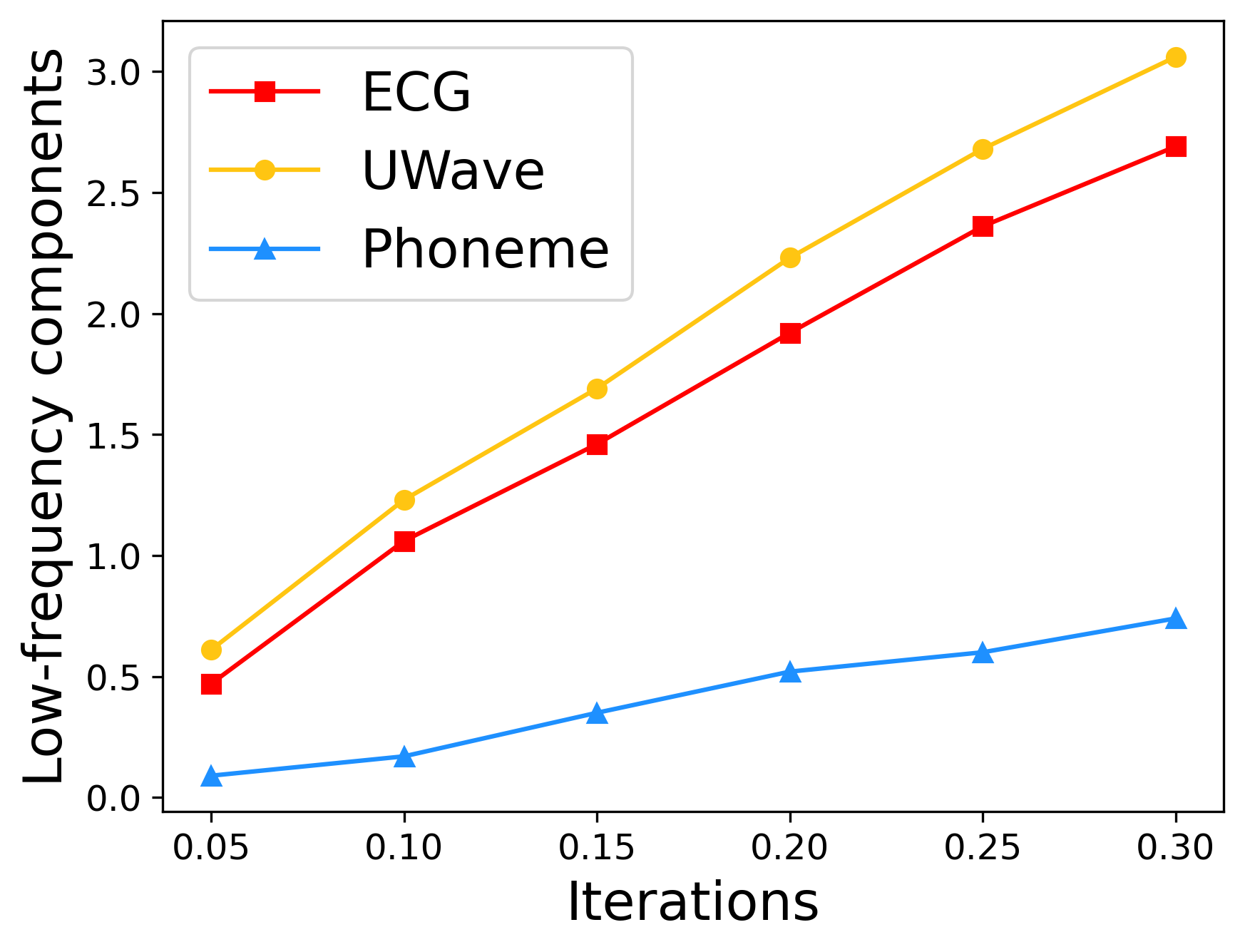}
  }
  \vspace{-0mm}
  \caption{Ablation study on budget on InceptionTime model}
  \vspace{-0mm}
  \label{Budget}
\end{figure}

We observe from Figure~\ref{Iteration} that the $L_2$ norm and DTW distance remain relatively stable concerning the number of iterations. While the LF component shows a gradual reduction with increasing iteration numbers. This is expected since more iterations construct more fluctuated perturbations, thereby reducing the LF component. In terms of the ASR, it gradually increases with the number of iterations. However, the ASR eventually converges, suggesting that there exists an inflection point in the relationship between the number of iterations and the performance of the adversarial attack. Beyond this inflection point, increasing the number of iterations yields diminishing improvement on ASR, but adds a considerable load on computation. The same phenomenon can be observed with respect to the perturbation budget in Figure~\ref{Budget}. Therefore, we choose the budget in the inflection point at around 0.25 which ensures both a satisfactory ASR and a high degree of imperceptibility for the adversarial samples.

\subsection{Case Study}

We further visualize the adversarial time series to show some qualitative results. We observe from Figure~\ref{Examples} that SFAttack conducts local perturbation where the non-shapelet part variation is near zero, which significantly reduces the magnitude of the perturbation. Besides, our approach produces perturbation with more oscillations rather than nearly square waves from baselines like FGSM, thus less perceptible to HVS and more difficult to detect. The qualitative further demonstrates the good imperceptibility of our proposed approach.

%% file: contents/05_conclusion.tex
\section{Conclusion}

In this paper, we propose the Shapelet-based Frequency-domain Attack (SFAttack). Our approach focuses on local perturbations targeting time series shapelets, which are more discriminative for TSC. Additionally, we introduce a low-frequency constraint to ensure perturbations remain in high-frequency components, further enhancing imperceptibility. Our theoretical and empirical analyses demonstrated that SFAttack generates less perceptible perturbations compared to traditional methods.

%% file: contents/06_appendix.tex
\clearpage

\section{Appendix}

\textbf{Theorem 1.} \emph{Suppose we denote the gradient of the time series $x$ with respect to the loss function $L(\cdot)$ as ${\nabla}L(x)$. If we conduct perturbations on the time series shapelet, to achieve the optimal performance of increasing the classification loss function, the perturbation can be computed as follows:}

\vspace{-4mm}
\begin{align}
    x^{adv} = x+{\epsilon}\cdot{\nabla}L_{n}(s)
\end{align}
\vspace{-3mm}

\vspace{-4mm}
\begin{align}
    {\nabla}L(x) = {\nabla}L(s)+{\nabla}L(\hat{x})
    \label{Decompose}
\end{align}
\vspace{-3mm}

\noindent \emph{where ${\nabla}L(s)$ and ${\nabla}L(\hat{x})$ denotes the gradients in the interval of shapelet $s$ and the remaining areas, respectively. While ${\nabla}L_{n}(s)$ stands for the normalized ${\nabla}L(s)$.}

\vspace{2mm}

\emph{Proof.} According to Taylor's Formula, we have:

\vspace{-4mm}
\begin{align}
    L(x+\epsilon\cdot\Delta{x}) = L(x)+\nabla{L(x)}\cdot\epsilon\cdot\Delta{x}
\end{align}
\vspace{-3mm}

Since the original loss $L(x)$ is a constant term, we focus on maximizing the difference value to make the adversarial example more effective. In other words, to achieve a higher increase in the loss, we need to maximize the following term under the constraint of $L_2$ norm. Besides, since the local perturbation is on the time series shapelet, the perturbation value on the non-shapelet area should be zero. We denote the shapelet area start and end point as $i_1$ to $i_2$, then the problem can be formulated as: 

\vspace{-4mm}
\begin{align}
\max_{\Delta{x}\in{R^T}} \quad &\nabla{L(x)}\cdot\epsilon\cdot\Delta{s}\\
s.t. \quad &\|\Delta s\|_2 \leq 1\\
&\Delta{s}_{i}=\left\{
\begin{aligned}
\Delta{x_i} & , & i_1\leq i\leq{i_2}, \\
0 & , & otherwise.
\end{aligned}
\right.
\end{align}
\vspace{-2mm}

Let us ignore the constant coefficient $\epsilon$ and according to the Cauchy-Schwarz Inequality, we have the following results:

\vspace{-4mm}
\begin{align}
&\nabla{L(x)}\cdot\Delta{s}\\
= \quad &\sum_{i=0}^{T-1}\nabla{L(x)_i}\cdot{\Delta{s_i}}\\
= \quad &\sum_{i=i_1}^{i_2}\nabla{L(x)_i}\cdot{\Delta{x_i}}\\
\leq \quad &\sqrt{\sum_{i=i_1}^{i_2}(\nabla{L(x)_i})^2}\cdot\sqrt{\sum_{i=i_1}^{i_2}(\Delta{x_i})^2}\\
= \quad & \|\nabla{L(s)}\|_2\cdot{\|\Delta s\|_2}\\
\leq \quad & \|\nabla{L(s)}\|_2
\end{align}
\vspace{-3mm}

The equality holds if and only if the gradient $\nabla{L(s)}$ and the perturbation performed on the shapelet $\Delta{s}$ are parallel. In other words, the perturbation $\Delta{s}$ can be formulated as:

\vspace{-4mm}
\begin{align}
    {\nabla}L_{n}(s) = \frac{\nabla{L(s)}}{\|\nabla{L(s)}\|_2}
\end{align}
\vspace{-2mm}

In a word, under the constraint of local perturbation on time series shapelet $s$, the best attack effectiveness can be achieved when the perturbation is equal to ${\nabla}L_{n}(s)$.

\vspace{4mm}

\textbf{Theorem 2.} \emph{With the same constraint of $L_2$ norm that $||G_{s}(x)-x||_2\leq\epsilon$ and $||G_{n}(x)-x||_2\leq\epsilon$ and denote the loss function and its increment as $L(\cdot)$ and ${\Delta}L(\cdot)$, respectively. Then, $G_{s}(x)$ is more effective than $G_{n}(x)$:}

\vspace{-4mm}
\begin{align}
{\Delta}L(G_{s}(x))>{\Delta}L(G_{n}(x))
\end{align}
\vspace{-4mm}

\noindent \emph{where $G_{s}(x)$ and $G_{n}(x)$ denotes the attacks on the shapelet and the non-shapelet part of the time series.} 

\vspace{2mm}

\emph{Proof.} Time series shapelets are subsequences that best split time series data into classes. Without loss of generality, we assume the length of the shapelet part and non-shapelet part are the same. Even if the lengths are different, the perturbation will be scaled due to the $L_2$ norm constraint, thus having no impact on the results. Specifically, suppose the TSC model can be denoted as $f(\cdot)$, then the following equation holds:

\vspace{-4mm}
\begin{align}
f(G_s(x))_k>f(G_{n}(x))_k
\label{f(x)}
\end{align}
\vspace{-4mm}

\noindent where the subscript $k$ denotes the ground-truth class for $x$. Then let us consider the cross entropy (CE) loss function and its gradient can be computed using the Chain Rule, which is shown as follows:

\vspace{-4mm}
\begin{align}
L(f(x)) &= -\sum_{j=1}^{|C|}y_j\cdot{\log{f(x)_j}}\\
&= -{\log{f(x)_k}}
\end{align}
\vspace{-3mm}

\vspace{-4mm}
\begin{align}
\nabla{L(f(x))}=-\frac{\nabla{f(x)_k}}{f(x)_k}
\end{align}
\vspace{-2mm}

According to Equation.\eqref{f(x)}, the output of $G_s(x)$ is larger and more close to ground truth than $G_n(x)$. As the output of the model $f(x)$ becomes larger and closer to the ground truth (\ie $f(x)$ converges to 1), the gradient 
$\nabla{f(x)}$ decreases. Thus, the following inequality holds:

\vspace{-4mm}
\begin{align}
\nabla{L(f(G_s(x)))}>\nabla{L(f(G_n(x)))}
\end{align}
\vspace{-2mm}

For ease of representation, it can be rewritten as follows:

\vspace{-4mm}
\begin{align}
\nabla{L(G_s(x))}>\nabla{L(G_n(x))}
\label{L(G(x))}
\end{align}
\vspace{-2mm}

According to Taylor's Formula, we have:

\vspace{-4mm}
\begin{align}
    \Delta{L(G_s(x))} = \nabla{L(G_s(x))}\cdot{(G_s(x)-x)}
\end{align}
\vspace{-4mm}

\vspace{-4mm}
\begin{align}
    \Delta{L(G_n(x))} = \nabla{L(G_n(x))}\cdot{(G_n(x)-x)}
\end{align}
\vspace{-4mm}

The maximum values of both $L(G_s(x))$ and $L(G_n(x))$ are achieved when the gradient and the perturbation are parallel. According to Equation.\eqref{L(G(x))}, we have following result:

\vspace{-4mm}
\begin{align}
{\Delta}L(G_{s}(x))>{\Delta}L(G_{n}(x))
\end{align}
\vspace{-2mm}

\vspace{2mm}

\textbf{Theorem 3.} \emph{Denote the $G_{FGM}(x)$ and $G_{FGSM}(x)$ as the adversarial example generated by FGM and FGSM, respectively. Given the same constraint of $L_2$ norm that $||G_{FGM}(x)-x||_2\leq\epsilon$ and $||G_{FGSM}(x)-x||_2\leq\epsilon$ and denotes the low-frequency component of a time series as $LF(\cdot)$, $G_{FGM}(x)$ contains less low-frequency component than $G_{FGSM}(x)$:} 

\vspace{-4mm}
\begin{align}
    LF(G_{FGM}(x))<LF(G_{FGSM}(x))
\end{align}
\vspace{-3mm}

\emph{Proof.} According to the definition of FGM and FGSM, the perturbations produced by them are normalized by $L_2$ norm and $L_\infty$ can be formulated as:

\vspace{-4mm}
\begin{align}
    p_1(x)_i = \epsilon\cdot{\frac{\nabla{L(x)_i}}{\|\nabla{L(x)}\|_2}}
\end{align}
\vspace{-2mm}

\vspace{-4mm}
\begin{align}
    p_2(x)_i = \epsilon\cdot\frac{sgn(\nabla{L(x)_i})}{\sqrt{T}}
\end{align}
\vspace{-3mm}

\noindent where $p_1(x)$ and $p_2(x)$ denote the perturbations generated by FGM and FGSM, respectively, and the $sgn(\cdot)$ represents the sign function.

Then, we compute the frequency spectrum of $p_1$ and $p_2$ through the Fourier Transform:

\vspace{-4mm}
\begin{align}
    \hat{p}(n) = \sum_{j=0}^{T-1}{e^{-i\frac{2\pi}{T}nj}}\cdot{p(x)_j}
\end{align}
\vspace{-2mm}

\noindent where $\hat{p}(j)$ denotes the frequency component of frequency $j$. Since the most important low-frequency component is the direct-current component, which is the most perceptible through human eyes, we compute the $LF(G_{FGM}(x))$ and $LF(G_{FGSM}(x))$ as follows:

\vspace{-4mm}
\begin{align}
    LF(G_{FGM}(x)) = |\sum_{j=0}^{T-1}{p_1(x)_j}|
\end{align}
\vspace{-2mm}

\vspace{-4mm}
\begin{align}
    LF(G_{FGSM}(x)) = |\sum_{j=0}^{T-1}{p_2(x)_j}|
\end{align}
\vspace{-2mm}

Typically, the gradient $\nabla{L(x)}$ is consecutive, and considering the sign of $\nabla{L(x)}$ keeps unchanged. No matter whether the sign is positive or negative, according to Cauchy-Schwarz Inequality, we have the following result:

\vspace{-4mm}
\begin{align}
|\sum_{j=0}^{T-1}{p_1(x)_j}| = \quad& |\sum_{j=0}^{T-1}\epsilon\cdot{\frac{\nabla{L(x)_j}}{\|\nabla{L(x)}\|_2}}|\\
\leq \quad & \epsilon\cdot\sqrt{\sum_{j=0}^{T-1}(\frac{\nabla{L(x)_j}}{\|\nabla{L(x)}\|_2})^2}\cdot{\sqrt{T}}\\
= \quad & \epsilon\cdot\sqrt{T}\\
= \quad & |\sum_{j=0}^{T-1}\epsilon\cdot\frac{sgn(\nabla{L(x)_j})}{\sqrt{T}}|\\
= \quad & |\sum_{j=0}^{T-1}{p_2(x)_j}|
\end{align}
\vspace{-2mm}

As the gradient series is composed of these consecutive and sign-unchanged segments, we have:

\vspace{-4mm}
\begin{align}
    LF(G_{FGM}(x))<LF(G_{FGSM}(x))
\end{align}
\vspace{-2mm}

\vspace{2mm}

\textbf{Theorem 4.} \emph{Given the same constraint of $L_2$ norm that $||G_{FGM}(x)-x||_2\leq\epsilon$ and $||G_{FGSM}(x)-x||_2\leq\epsilon$ and denote the loss function of the TSC classifier $f$ and its increment as $L(\cdot)$ and ${\Delta}L(\cdot)$, respectively. Then, $G_{FGM}(x)$ is more effective than $G_{FGSM}(x)$:}

\vspace{-4mm}
\begin{align}
    {\Delta}L(G_{FGM}(x))>{\Delta}L(G_{FGSM}(x))
\end{align}
\vspace{-2mm}

\emph{Proof.} According to Taylor's Formula, we have:

\vspace{-4mm}
\begin{align}
    {\Delta}L(G_{FGM}(x)) = \nabla{L(x)}\cdot\epsilon\cdot(G_{FGM}(x)-x)
\end{align}
\vspace{-4mm}

\vspace{-4mm}
\begin{align}
    {\Delta}L(G_{FGSM}(x)) = \nabla{L(x)}\cdot\epsilon\cdot(G_{FGSM}(x)-x)
\end{align}
\vspace{-4mm}

For ease of presentation, we ignore the coefficient $\epsilon$ and rewrite them as follows:

\vspace{-4mm}
\begin{align}
    {\Delta}L(G_{FGM}(x)) = \nabla{L(x)}\cdot\Delta{x_1}
\end{align}
\vspace{-4mm}

\vspace{-4mm}
\begin{align}
    {\Delta}L(G_{FGSM}(x)) = \nabla{L(x)}\cdot\Delta{x_2}
\end{align}
\vspace{-4mm}

According to the previous results, $\Delta{L(x)}$ achieves the maximum value when $\Delta{x}$ is parallel to the gradient $\nabla{L(x)}$. In other words, the optimal value of $\Delta{x_1}$ is:

\vspace{-4mm}
\begin{align}
    \Delta{x_1^\ast} = \frac{\nabla{L(x)}}{\|\nabla{L(x)}\|_2}
\end{align}
\vspace{-3mm}

As for $\Delta{x_2}$, its magnitudes at each timestamp are bounded:

\vspace{-4mm}
\begin{align}
    |\Delta{x_2}(n)| = 1
\end{align}
\vspace{-3mm}

The optimal value may not be achieved as the direction may not be aligned with $\nabla{L(x)}$ under the magnitude constraint. Thus, we have:

\vspace{-4mm}
\begin{align}
    {\Delta}L(G_{FGM}(x))>{\Delta}L(G_{FGSM}(x))
\end{align}
\vspace{-2mm}